\documentstyle[12pt,aasms4]{article}

\received{}
\accepted{}
\journalid{}{}
\articleid{}{}

\slugcomment{accepted by ApJ Letters}

\lefthead{Martin et al.}
\righthead{Discovery of a Very Low-Mass Binary with HST/NICMOS}

\begin{document}

\title{Discovery of a Very Low-Mass Binary with HST/NICMOS}

\author{E.L. Mart\'\i n and G. Basri}
\affil{Astronomy Department, University of California,
    Berkeley, CA 94720}


\author{W. Brandner}
\affil{Jet Propulsion Laboratory/IPAC, Mail Code 100-22, Pasadena, CA 91125}


\author{J. Bouvier}
\affil{Observatoire de Grenoble, B.P.53, 
F-38041 Grenoble Cedex 9, France}


\author{M. R. Zapatero Osorio and R. Rebolo}
\affil{Instituto de Astrof\'\i sica de Canarias, 38200 La Laguna, Spain}

\author{J. Stauffer}
\affil{Smithsonian Astrophysical Observatory, 60 Garden St., Cambridge, 
MA 02138}

\author{F. Allard and I. Baraffe}
\affil{CRAL, Ecole Normale Superieure, 46 Alee d'Italie, Lyon, 
69364 France}

\and
\author{S. T. Hodgkin}
\affil{Astronomy Group, Department of Physics and Astronomy, Leicester 
University, University Road, Leicester, LEI 7RH, England, UK}

\centerline{contact e-mail address: ege@popsicle.berkeley.edu} 

\begin{abstract}

Hubble Space Telescope NICMOS observations are presented of six brown dwarf
candidates in the Pleiades open cluster.  One of them, namely CFHT-Pl-18, is
clearly resolved as a binary with an angular separation of 0".33.  
The very low density of contaminating background stars in our
images and the photometry of the components 
support that this system is a physical binary rather
than a chance projection. All the available photometric and
spectroscopic data indicate that the CFHT-Pl-18 system is likely a 
member of the Pleiades cluster, but a final confirmation will have 
to wait until lithium can be detected. Assuming cluster membership,  
we compare our NICMOS photometry 
with evolutionary models, and find that the inclusion of the effects
of dust grains is necessary for fitting the data. 
We estimate that the masses of the
components are about 0.045~M$_\odot$ and 0.035~M$_\odot$. 
The binary system has a projected separation of 42~AU (for a distance of 
125~pc) that is common among stellar binaries.

\end{abstract}


\section{Introduction}

Brown dwarfs (BDs; substellar objects with masses lower than about
0.075~M$_\odot$) cool down with increasing age to very low temperatures and
faint luminosities, becoming very difficult to detect.  Ten cool Pleiades members have been shown to
have strong resonance Li\,{\sc i} lines that confirm their BD status 
and provide a nuclear age of 120~Myr for the cluster (Basri,
Marcy \& Graham 1996; Rebolo et al.  1996; Mart\'\i n et al.  1998a;
Stauffer, Schultz \& Kirkpatrick 1998).  Thus, the substellar borderline and
the cluster sequence into the BD realm is now well established.  The
substellar limit in the Pleiades is located at I$\sim$17.8 and spectral type
M6--M6.5 (Mart\'\i n, Rebolo \& Zapatero Osorio 1996; Stauffer et al. 1998).

Recent deep CCD surveys in the Pleiades have been successful in revealing a
numerous population of BD candidates (Zapatero-Osorio et al. 1997a;
Bouvier et al.  1998; Festin 1998).  The density of BD candidates uncovered by
these surveys indicates that the mass function (MF) of the Pleiades does not
turnover in the stellar domain.  Mart\'\i n, Zapatero Osorio \& Rebolo (1998)
and Bouvier et al.  (1998) have shown that the Pleiades MF can be approximated
to a power law M$^{-\alpha}$ with $\alpha$ in the range 0.5 to 1.2 for masses
between 0.4~M$_\odot$ and 0.04~M$_\odot$.

The substellar MF of the Pleiades cluster has not been corrected for binaries
because nothing is known about the multiplicity of brown dwarfs.  The effect
of binarity is to increase the number of BDs, and thus the MF slope $\alpha$
should be revised upward if there are many BD companions.  Recently it has
been shown that the Pleiades object PPl~15 is a double-lined spectroscopic
binary (Basri \& Mart\'\i n 1998) composed of two brown dwarfs.  Since very
few BDs have been checked for radial velocity variations, this suggests a
high binary fraction among BDs.  With the aim of finding more BD
binaries, we selected a sample of 30 targets with high likelihood of being
Pleiades members with masses between 0.09~M$_\odot$ and 0.04~M$_\odot$ for 
an imaging survey with the NICMOS camera on the Hubble Space Telescope 
(HST).  This
paper reports on the observations of the first 6 objects of our program.  
One of them has been clearly resolved into two components and we argue 
that it probably constitutes the first resolved BD--BD binary.

\section{Observations, Results and Discussion}

The HST observations of our first 6 
targets were obtained between 1998 February 17 and 1998 March 31. 
The objects were selected randomly among our total list  
of 30 targets that were 
chosen to be representative of all known Pleiades BD candidates. 
Our sample includes `Calar' and `Teide' objects from Zapatero Osorio et al. 
(1997b) and Mart\'\i n et al. (1998a); `CFHT' objects from Bouvier et
al.  (1998); `HHJ' objects from  Hambly, Hawkins \& Jameson
(1993); `MHO' objects from Stauffer et al.  (1998b); and `Roque' objects
from Zapatero Osorio et al.  (1997a). 

We used the NICMOS camera 1 (NIC1) in multiple-accumulate mode with filters
F110M, F145M, and F165M (Thompson et al. 1998). 
 The integration times were 896~s in F110M and 768~s in the other 
two filters.  The
limiting sensitivities are 0.012~mJy in F110M, 0.017~mJy in F145M and
0.016~mJy in F165M (corresponding to $J$=20.0 and $H$=19.4).  
They were derived based on the assumption
that we would detect a source if the counts in the central pixel of its PSF
are at least 3 times the sigma of the background. 
Each target was observed during only one orbit (visibility 52 min).

In Figure 1 we show the NIC1 images of five targets.  All of them are
consistent with being unresolved single objects except for CFHT-Pl-18, which
is clearly resolved into two components.  The separation between them is
0.334$\pm$0.001 arcsec and the position angle is 351.3$\pm$0.2 degrees.  At
the distance of the Pleiades cluster (125 pc) the observed angular separation
corresponds to a projected binary axis distance of 41.75 AU.  Flux values for
all the targets were computed based on the header keyword PHOTFNU and are
given in Table~1.  These values should be used with caution because the values
for PHOTFNU are valid for sources with a constant flux per unit wavelength
across the band pass, which might not be the case for our sources.  Other
limiting factors in the accuracy of the fluxes are that we used model PSF for
fitting the data that did not provide a perfect match to the observed PSF, 
and uncertainties in the NICMOS darks and resulting spatial variations 
in the background.  We
used on-orbits darks (as opposed to the model darks used in the standard
NICMOS pipeline) to improve the photometric accuracy, but it was still not
perfect.

We obtained near-IR photometry of CFHT-Pl-18 at the 3.6~m CFHT telescope on
13-15 January 1998.  Using zero-points from UKIRT faint standards we obtained
the following magnitudes:  J=15.95$\pm$0.02; H=15.23$\pm$0.04 and
K=14.80$\pm$0.07.   After discovering
the binary nature of CFHT-Pl-18 in the NIC1 data, we applied for service
observations at the Keck observatory.  On 1998 August 8, an LRIS spectrum was
obtained at Keck~II by the staff of the observatory.  The
1200/7500 grating was used with a 1$''$ slit.  One exposure of 1800~s
was obtained.  The CCD frame was reduced and wavelength calibrated using
standard IRAF routines.  The spectral dispersion, FWHM resolution and range
recorded are:  1.26~\AA , 2.4~\AA ~ and 664.6--793.3~nm, respectively.

The $RIJHK$ photometric measurements of CFHT-Pl-18 support the membership to
the Pleiades cluster because they are similar to the benchmark BDs Calar~3
and Teide~1 (Zapatero-Osorio et al. 1997c).  
Additional evidence for cluster membership is provided by the
LRIS spectrum.  In Figure~2 we compare the spectra of CFHT-Pl-18 and 
Teide~1 (Rebolo et al. 1995, 1996). We do not find any significant 
difference between them.  We obtained a
heliocentric radial velocity for CFHT-Pl-18 of 2.3$\pm$10.5~km s$^{-1}$ 
by cross-correlation with an LRIS spectrum of VB~10 
(V$_h$=35~km s$^{-1}$) obtained in another run.  
This radial velocity is fully consistent with cluster membership, but does 
not rule out that CFHT-Pl-18 could be a field young disk star.  
Our spectrum of
CFHT-Pl-18 is unfortunately too noisy to attempt a detection of the lithium 
resonance line at 670.8~nm. We intend to obtain higher quality spectra soon.

Zapatero Osorio et al.  (1997b) discussed the different kinds of objects that
could contaminate the Pleiades photometric sequence and found that only 
very-low mass (VLM) foreground stars could be important.  We have estimated the
probability that CFHT-Pl-18 could be a field VLM star.  The Keck spectrum
yields a spectral type of M8 using the indices defined by 
Mart\'\i n et al. (1996). 
If it were a main-sequence M8 dwarf, it would have
to be at a distance of about 90~pc.  The local density of M8 dwarfs is about
0.0024 pc$^{-3}$ (Kirkpatrick et al. 1994).  The number of expected M8 dwarfs
in a distance range between 80~pc and 100~pc in the total area of the CFHT CCD
survey (2.5 deg$^2$; Bouvier et al. 1998) is $\sim$0.8.  Since there are 6 BD
candidates in the CFHT survey with magnitudes in the range $I$=18.6 and 19.0,
the probability that CFHT-Pl-18 is a contaminating foreground star is
$\sim$15\% .  Thus, it is much more likely that CFHT-Pl-18 is a 
young BD binary system rather than an old field VLM stellar binary, and 
our measured radial velocity is a positive consistency check of Pleiades 
membership. Eventually it would be good to have proper motion confirmation.

CFHT-Pl-18 is likely to be a physical binary rather than a chance coincidence
in the sky of two unrelated objects because of three reasons: The
density of contaminating stars in our NIC1 images is very low (Figure~1).  
The fainter component of CFHT-Pl-18 is  redder than the primary and 
probably  has stronger water absorption.  
The F145N filter is strongly affected by water vapour.  The B/A
ratio in this filter is lower (0.480$\pm$0.011) than in the F110M filter 
(0.488$\pm$0.010), indicating
stronger water absorption in the B component than in the A component,
consistent with being cooler.

In order to derive the luminosities and masses of the CFHT-Pl-18 
components we assume that it belongs to the Pleiades.  This
fixes the age, distance and metallicity at 120~Myr, 125~pc and solar,
respectively. In Figure~3, we present a comparison of our data with 
state of the art models. 
The ``Dusty'' models include dust grains in the equation
of state and the opacities, and the ``Cond'' models include grains in the
equation of state but not in the opacities.
The NextGen models (Allard et al.  1997; Baraffe et
al.  1998) do not have any grains and they are systematically redder than the
Dusty and Cond models, even in the stellar domain, because they were computed
with a different water molecule data set.  The NextGen models provide a good
fit to the data for the higher masses (0.08--0.06~M$_\odot$), but the Cond
models fit better the lower masses (0.06--0.03~M$_\odot$).  
Our models should be considered
preliminary because they do not do a very good job reproducing the $H$
bandpasses of very cool dwarf spectra (Allard et al. 1998). 
We plan in the future to make a
detailed comparison between models and the NICMOS data of our whole 
sample. Presently, we
note that the low-mass Pleiades BDs are likely to be dusty, and that the
masses estimated for them depend on the details of the models. 
Taking into account the observational error bars and the theoretical
uncertainties, we estimate masses of M/M$_\odot$=0.046$\pm$0.006 and
M/M$_\odot$=0.035$\pm$0.007 for CFHT-Pl-18~A and B, respectively.

Bouvier et al.  (1997) have performed a search for close multiple systems to
G- and K-type Pleiades members. 
They found companions with separations in the range
0.08$''$--6.9$''$ and a contrast in the H-band up to 5 magnitudes.  They
derived a binary frequency of 28$\pm$4\% in the orbital period range from 4.2
to 7.1 log(days) after correction for incompleteness.  Such binary frequency
is similar to that of nearby field G-type dwarfs (27\%).  The separation of
the components of the CFHT-Pl-18 system lies around the maximum of the
distribution of binary separation among different kinds of low-mass stars.  If
the multiplicity frequency of BDs is similar than for solar-type stars (a
hypothesis that will be tested with our observations) we expect to find at
most 8 binaries among our 30 primaries, depending on the mass-ratio
distribution of brown dwarf binaries.  So far we have found 1 binary out of 6
candidate primaries (16\% detection rate), consistent with the field dwarf
results.
The study of the multiplicity in our sample is important for constructing the
MF below the substellar limit.  The Pleiades cluster is the first site where a
sufficient number of BDs can be studied in order to derive a substellar mass
function.  The recent La Palma and CFHT surveys strongly indicate that the MF
rises below the substellar limit (Zapatero Osorio et al.  1997a; Mart\'\i n et
al.  1998b; Bouvier et al.  1998).  The effect of unresolved binaries is an
important uncertainty in the computation of the MF (Kroupa 1995).  The fact
that PPl~15 and CFHT-Pl-18 are binaries indicates that BD companions 
suggests that binary BDs may be frequent. We plan to discuss the 
binary statistics in a future paper when our HST program has been completed. 

The binary CFHT-Pl-18 may have formed via the gravitational collapse and late
fragmentation of a very low-mass molecular core.  
Since the opacity limit for gravitational collapse is thought
to be as low as $\sim$0.01~M$_\odot$ (Kanjilal \& Basu 1992), 
there is no reason why the BDs should form in a manner different from stars  
unless such fragmentation does not proceed to completion.
This latter point has been an important question to be addressed by the
discovery of brown dwarfs.  Now that many free-floating BDs have been found,
hierarchical fragmentation might be considered a viable hypothesis, and the
discovery of this binary suggests that the process of binary 
formation could be similar in the substellar and stellar mass ranges.

\acknowledgments

{\it Acknowledgments}: 
This research is based on data collected at the 
Hubble Space Telescope, the Canada-France-Hawaii telescope  
and the Keck~II telescope. 
We are grateful to the staff of the Keck observatory for 
carrying out LRIS service observations of CFHT-Pl-18. 
We thank G. Chabrier for his help in developing the models used in this paper.  
EM acknowledges the support from the F.P.I. program 
of the Spanish Ministry of Education and Culture. 
GB acknowledges the support of NSF through grant AST96-18439. 
FA was supported by NASA through grants 110-96LTSA and NAG5-3435.
Funding for this publication was provided by NASA through Proposal 
GO-7899 submitted to the Space Telescope Science Institute, which is operated 
by the Association of UNiversities for Research in Astronomy, Inc., under
NASA contract NAS5-26555. 


\clearpage

\centerline{\bf Figure Captions:}

\figcaption[brandner.ps]{\label{fig1} Mosaic of reduced HST/NIC1 images.}

\figcaption[spectra.eps]{\label{fig2} The optical spectrum of CFHT-Pl-18 
compared with Teide~1 (Rebolo et al. 1996). Both spectra have similar 
spectral resolution and have been smoothed with a boxcar of 3 pixels. 
The spectrum of CFHT-Pl-18 has been normalized in the same 
region ($\sim$690~nm) 
as the spectrum of Teide1, and shifted upward by 5 units.}

\figcaption[colmagnic.eps]{\label{fig3} A color-magnitude with the NICMOS 
filters. The two components of CFHT-Pl-18 are marked with filled 
hexagons. The other program objects are marked with empty circles. 
The solid line is the 120 Myr isochrone for NextGen model atmospheres 
(Allard \& Hauschildt 1995). 
The dotted and dashed lines are isochrones for the same age but  
including dust effects and a different water line list 
(see text for details). The asterisks on the isochrones 
denote the position of models with masses 0.075 and 0.040~M$_\odot$.}


\begin{deluxetable}{lcccc}
\footnotesize
\tablecaption{\label{tab1} HST data for our program objects}
\tablewidth{0pt}
\tablehead{
\colhead{Name}  &
\colhead{F110M} &
\colhead{F145M}  &
\colhead{F165M}  & 
\colhead{Separation}   
\nl   }                            
\startdata 
CFHT-Pl-18 A & 0.321$\pm$0.020 & 0.382$\pm$0.011 & 0.519$\pm$0.018 & 
0.334$\pm$0.001 \nl
CFHT-Pl-18 B & 0.157$\pm$0.020 & 0.183$\pm$0.011 & 0.266$\pm$0.018 &  \nl
CFHT-Pl-20   & 0.263$\pm$0.012 & 0.353$\pm$0.010 & 0.459$\pm$0.008 & 
$\le$0.08 \nl
HHJ8         & 1.193$\pm$0.022 & 1.452$\pm$0.013 & 1.800$\pm$0.017 & 
$\le$0.08 \nl
MHO6         & 0.800$\pm$0.029 & 0.960$\pm$0.018 & 1.234$\pm$0.023 & 
$\le$0.08 \nl
Roque 11     & 0.428$\pm$0.012 & 0.505$\pm$0.009 & 0.706$\pm$0.013 & 
$\le$0.08 \nl
Roque 12     & 0.450$\pm$0.021 & 0.566$\pm$0.012 & 0.767$\pm$0.014 & 
$\le$0.08 \nl
\enddata
\tablenotetext{}{The fluxes are given in mJy and the separations 
in arcsec. For unresolved objects an upper limit is given for companions 
up to 3 magnitudes fainter than the primary. The error bars are 
1~$\sigma$ standard deviations.}
\end{deluxetable}


\plotone{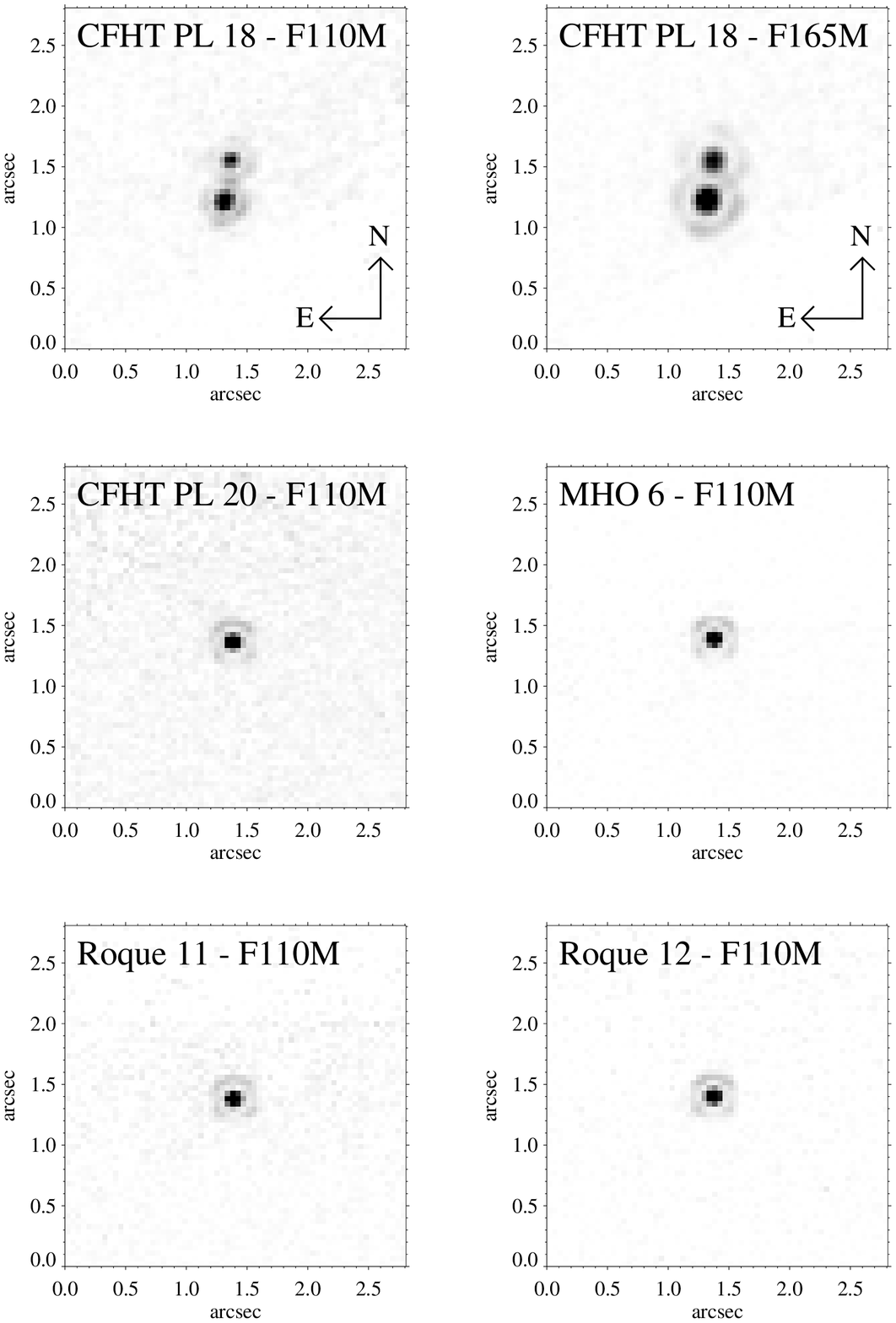}

\plotone{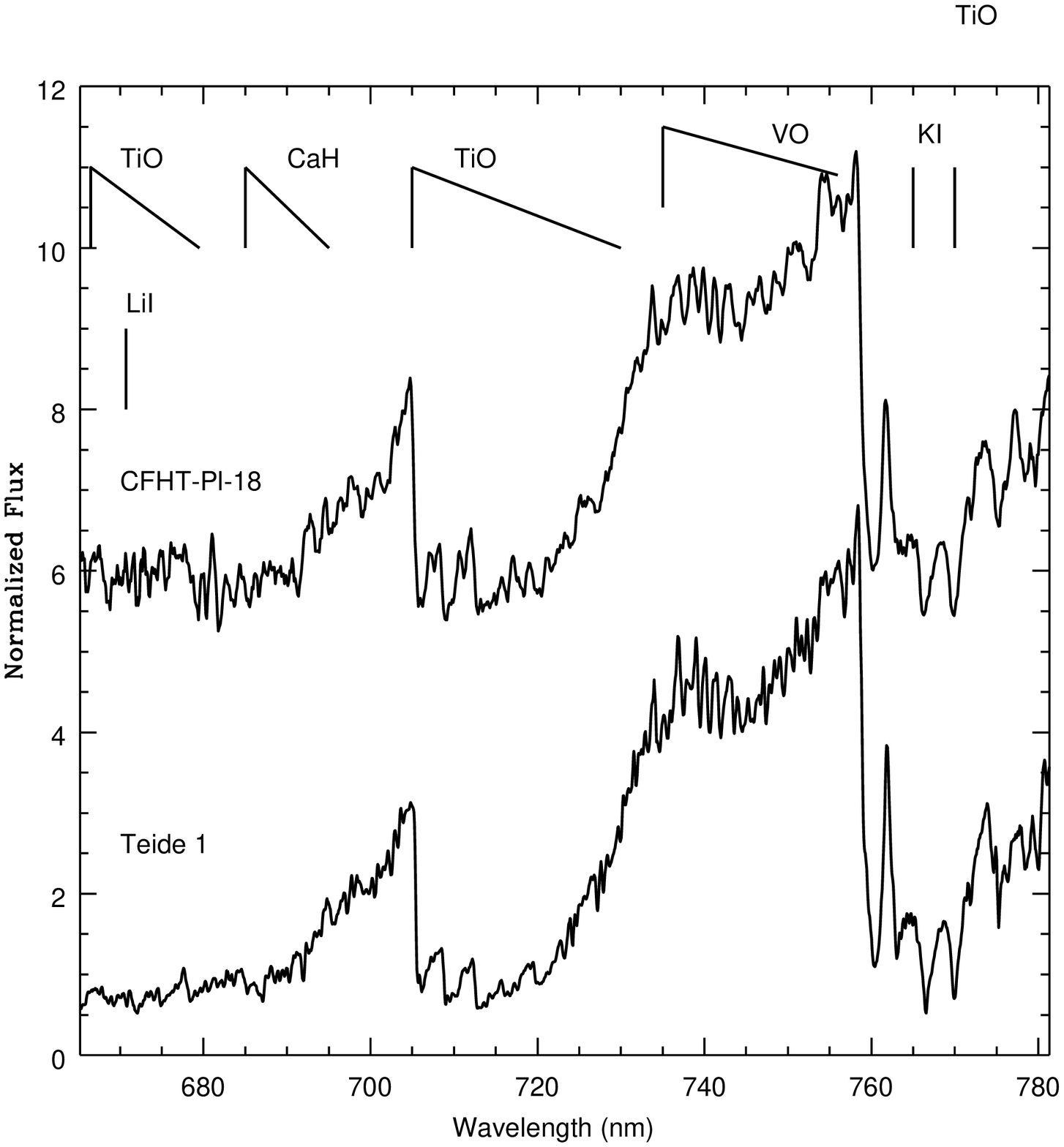}

\plotone{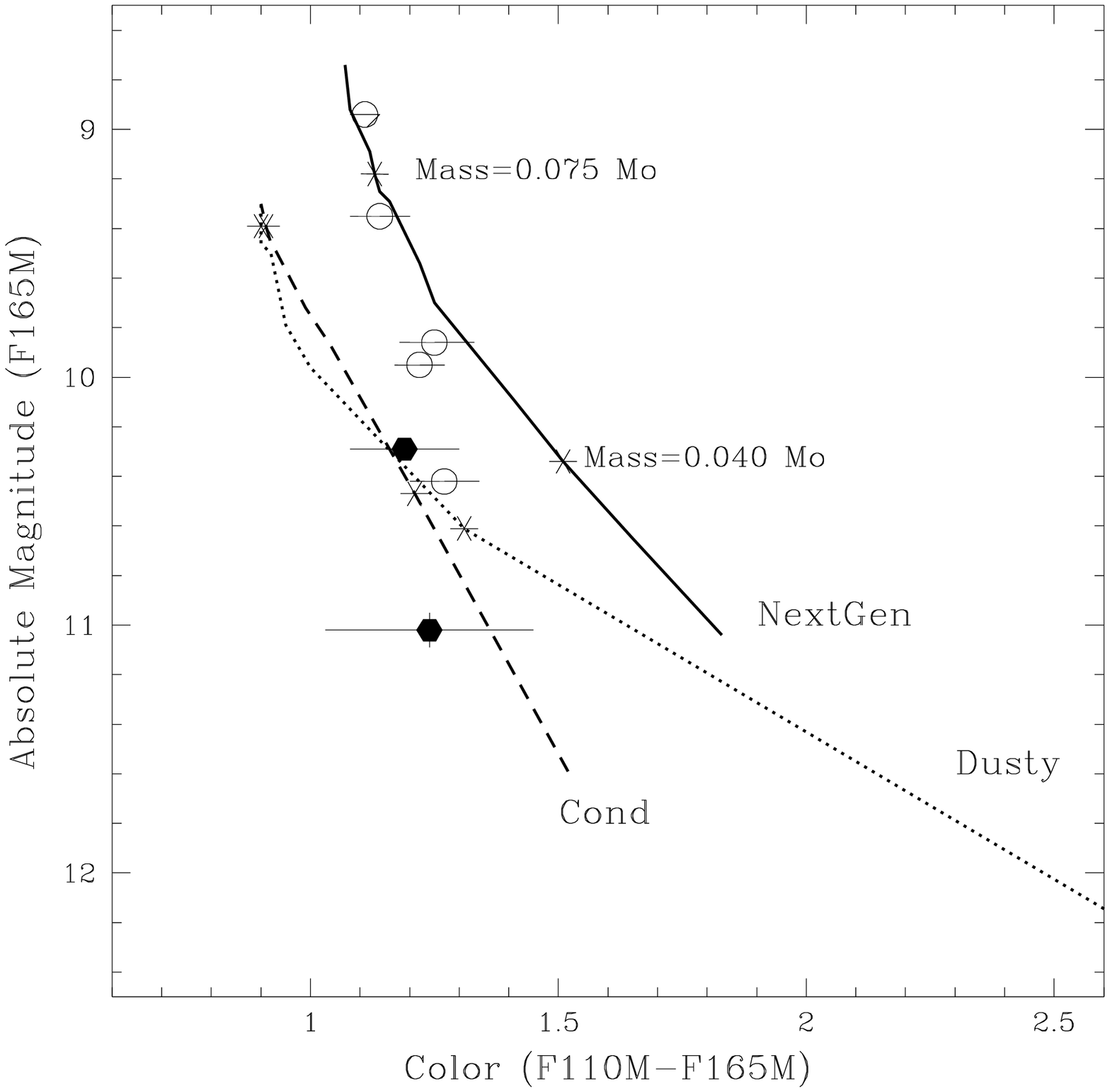}

\end{document}